# Quantum key distribution for 10 Gb/s dense wavelength division multiplexing networks

K. A. Patel,[1,2] J. F. Dynes,[1] M. Lucamarini,[1] I. Choi,[1] A. W. Sharpe,[1] Z. L. Yuan,[1] R. V. Penty[2] and A. J. Shields[1]

[1]*Toshiba Research Europe Limited, Cambridge Research Laboratory, 208 Cambridge Science Park, Milton Road, Cambridge, CB4 0GZ, United Kingdom*

[2]*Cambridge University Engineering Department, 9 J J Thomson Avenue, Cambridge, CB3 0FA, United Kingdom*

**Abstract**: We demonstrate quantum key distribution (QKD) with bidirectional 10 Gb/s classical data channels in a single fiber using dense wavelength division multiplexing. Record secure key rates of 2.38 Mbps and fiber distances up to 70 km are achieved. Data channels are simultaneously monitored for error-free operation. The robustness of QKD is further demonstrated with a secure key rate of 445 kbps over 25 km, obtained in the presence of data lasers launching conventional 0 dBm power. We discuss the fundamental limit for the QKD performance in the multiplexing environment.

Quantum key distribution (QKD)[1] is revolutionizing the way of distributing secret cryptographic keys for information protection. Unlike its classical counterparts, it allows secure transmission of information with quantifiable security[2-4] and detection of an eavesdropper. After numerous experiments in the laboratory[4-7] as well as in field,[8,9] the technology has matured for deployment on dark fiber with secure key rates exceeding 1 Mb/s. Nevertheless, for wider adoption, QKD must adapt to existing fiber infrastructures, which are populated with classical data communications.

Quantum and classical data signals can be combined onto a single fiber using wavelength division multiplexing. In an attempt to reduce signal deterioration, prior work often placed the quantum channel in 1300 nm band while the data channels are placed in 1550 nm band.[10-13] Using a combination of reduced data laser power and spectral filtering, multiplexing distances have reached up to 50 km with the secure bit rates limited to a few kb/s.[14,15] Recently, we have employed intrinsic temporal filtering provided by GHz-gated InGaAs single photon detectors and have dramatically improved the QKD performance in data-populated fiber. With coarse wavelength division multiplexing (CWDM) and bi-directional 1 Gb/s data communication, the QKD system was able to distribute secure keys at a rate of 500 kb/s over 50 km fiber.[16]

The next step is to adopt dense wavelength division multiplexing (DWDM) with 10 Gb/s data, but this is far more challenging. Firstly, a 10 Gb/s data channel requires 10 times higher laser power than 1 Gb/s, thus producing more photon scattering (Raman noise) into the quantum channel. Secondly, denser wavelength population will naturally accommodate a higher number of data channels. Thus, the Raman noise will increase further with each added data channel.

There have been various attempts to multiplex QKD with 10 Gb/s data using DWDM, but the resulting bit rates are low and distances limited.[17-21] Despite these advances, further studies are required to establish the true potential of QKD in DWDM 10 Gb/s networks. Higher secure key rates are needed to serve multiple users[22] and the communication distance must be extended for wider network coverage. For example, the IEEE standard[23] for extended reach has defined the 10 Gb/s communication distance of 40 km.

Here, we demonstrate the coexistence of QKD and error free 10 Gb/s data for various provisioning of the data channels. With reduced data laser power, we achieve a record fiber distance of 70 km and secure bit rates up to 2.38 Mb/s. With conventional data laser power, providing a total launch power of +3 dBm, we achieve a fiber distance of 25 km with a secure bit rate of 445 kb/s. All secure rates reported are at least two orders of magnitude higher than previous rates for equivalent distances and configurations. We also discuss the fundamental limit for the QKD performance in the multiplexing environment.

Figure 1 shows the experimental setup. Both quantum and classical channels are placed on an International Telecommunication Union (ITU) grid spaced by 100 GHz using off-the-shelf DWDM components. All the signals are multiplexed using an 8-channel DWDM module at each end of the communication channel realised with a dispersion shifted fiber. Each DWDM module covers the ITU grid channels 30 to 37 with characterized channel losses ranging from 0.72 to 1.54 dB. Isolation of non-adjacent channels varies from 77 to 90 dB, while that for adjacent channels can be significantly worse, measured to be between 43 and 84 dB. However, poorer isolation can be corrected by additional spectral filtering so that it is still possible, although unnecessary, to place data signals right next to the quantum channel. We assign channel 36 (1548.52 nm) to QKD and channel 34 (1550.12 nm) to the clock signal. 10 Gb/s channels are always launched in pairs to offer bidirectional communications. For the first pair, we assign Channel 32 (1551.72 nm) to co-propagate with the quantum signal, while Channel 33 (1550.92 nm) is counter-propagating. The clock synchronisation signal is transmitted optically at 10 MHz while extra small form pluggable (XFP) modules are used to provide 10 Gb/s data channels. The QKD system implements the T12 protocol,[4] which is an efficient version of the decoy-state[24,25] BB84 protocol featuring security in the finite-size scenario. In the quantum receiver, self-differencing InGaAs detectors[26] are used for single photon detection and also act as a temporal filter for noise rejection.[16] The secure key rates reported in this paper are obtained from key sessions of 20 minutes with a key failure probability of $\varepsilon=10^{-10}$.[4]

Raman scattering from data lasers in optical fiber can severely deteriorate QKD performance.[10-16] Owing to its wide spectral coverage exceeding 200 nm, most of the Raman photons are rejected spectrally by the DWDM module, which has a measured passband width of 0.56 nm. The Raman dip in the wavelength spectrum of data channels is naturally exploited here to further reduce the noise contribution to the quantum channel. Temporal filtering of the self-differencing detectors contributes also a 10 dB reduction in the Raman noise.[16] Further noise reduction can be achieved through a careful control of the data laser power. As shown in the inset of Fig. 2, only -27.0 dBm receiving power is required to achieve error-free 10Gb/s communication (bit error rate $\leq 10^{-12}$). Hence, transmitting at full power, typically at 0 dBm, is usually unnecessary. Therefore, in the first set of experiments we adapt the data laser power for each fiber distance to reduce the Raman noise. For example, for 50 km fiber, a data laser launching at -14.0 dBm is sufficient for error-free 10 Gb/s communication.

With adapted data laser power, we operate QKD over fiber distances of 35, 50, 65 and 70 km. The experimental result is plotted in Fig. 2 as a function of fiber link loss. For clarity only the quantum bit error rate (QBER) in the Z-basis (majority) is presented although the QBER is separately measured in both X and Z bases, as required by the T12 protocol.[4] The QBER increases with the link loss (fiber distance), illustrating the deterioration in the signal to noise ratio. Such deterioration is due to both the decreasing signal photon arrival rate and the increasing Raman noise. Higher link losses require higher data laser power and therefore result in more Raman noise. The secure key rate decreases exponentially at small link losses, but falls sharply for link losses greater than 14 dB. The sharp fall is due to the privacy amplification cost arising from an increase in QBER. The secure key rates are 2.38 Mbps, 1.17 Mbps, 177 kbps and 52 kbps for fiber distances of 35, 50, 65 and 70 km, respectively. Using experimental parameters, we have reproduced the obtained results by a numerical simulation (solid lines, Fig. 2).

We now study the ultimate data bandwidth that can be incorporated into a single fiber with QKD. Here, we keep the fiber distance constant at 50 km. As before, channels 32 (1551.72 nm) and 33 (1550.92 nm) carry 10 Gb/s data and co- and counter-propagate with the quantum signal, respectively. After carefully establishing the launching powers required for error-free communication, we raise the power of the data lasers to experimentally simulate communication at higher bandwidths. In this experimental simulation, we consider only a limited number of channels, with similar wavelengths and Raman scattering coefficients. In this case, both the Raman noise and communication bandwidth are linear with optical power. Moreover, we neglect any non-linear effects as we were unable to observe them in the power range of interest here.

Figure 3 shows the measured secure bit rate and QBER as a function of the simulated data bandwidths. We also measure a reference set with data lasers switched off. As data bandwidth increases, the QKD performance deteriorates gradually. We infer a maximum of fourteen 10 Gb/s channels can be integrated into a single fiber with QKD, providing an aggregated bandwidth of 140 Gb/s. This result suggests that QKD is sufficiently robust in metro networks carrying greater than 100 Gb/s data traffic. Coexisting with 100 Gb/s data, the measured secure rate is 342 kb/s, which remains approximately 30% of the value obtained without data channels. Theoretical simulation, shown as solid lines in Fig. 3, is in excellent agreement with experimental results.

In addition to the experimentally simulated data bandwidth, we have also experimentally performed QKD in presence of four 10 Gb/s channels (data bandwidth = 40 Gb/s). Two additional channels are placed at Channel 30 (1553.32 nm) and 31 (1552.52 nm) in co and counter-propagating directions respectively with respect to the QKD channel. We have obtained a secure key rate of 842 kb/s, as shown in Fig. 3. This result agrees with both experimental and theoretical simulations.

So far, we have considered a situation in which the power of the data lasers can be controlled. In the scenario where the data lasers cannot be attenuated, QKD has to operate in a high data laser power environment, for example 0 dBm. To test QKD's adaptability, we launch bidirectional 10 Gb/s lasers at 0 dBm at Channels 31 (1552.52 nm) and 30 (1553.33nm) which gives a combined launch power of 3 dBm. A standard off-the-shelf 25 GHz filter is inserted just before the quantum receiver to improve Raman noise rejection. The insertion loss of the filter is 5 dB. Figure 4 shows continuous operation of this high power data multiplexed QKD system over 12 hours and a fiber distance of 25 km. The QBER in the Z and X bases are 6.77 % and 5.99% respectively. Such error rates are sufficiently low to allow secure key generation. In fact, the secure key rate, averaged over 12 hours, is 445 kb/s.

The inset of Fig. 4 shows the QKD secure key rate over 25 km as a function of the data launch powers. It can be seen that a secure key rate of 100 kb/s is still obtainable with a combined data laser launch power of +5 dBm. The measured data (symbols) agrees with the theoretical simulation (line). These results illustrate the applicability of QKD to fiber networks with conventional data laser launch powers.

We now comment on the general aspects of QKD/data multiplexing. The main limitation for QKD in a DWDM network arises from Raman noise, which can be minimised either by reducing the data laser power or by narrowing the time and frequency filters. The former case has been investigated in the first part of this work, showing that a distance of at least 70 km is achievable. For the latter case, we consider the physical constraint imposed by the time-bandwidth product (TBP) when filtering simultaneously in both time and frequency.

If $\Delta t$ and $\Delta \nu$ are the full-width at the half maximum (FWHM) of a Gaussian pulse in the time and frequency domains, respectively, their TBP is lower bounded by:[27]

$$\Delta \nu \times \Delta t \geq \frac{2 \ln 2}{\pi} \cong 0.44. \qquad (1)$$

In our experiments, temporal filtering is determined by timing characteristics of single-photon detectors, which have an active timing window of $\Delta t = 100$ ps.[16] In the spectral domain, we used two sets of filters: (1) 100 GHz DWDM filter, with measured FWHM of $\Delta \nu_1 = 70$ GHz, and (2) off-the-shelf 25 GHz filter, with a measured FWHM of $\Delta \nu_2 = 15$ GHz. The corresponding TBP's for these two cases are 7 and 1.5, respectively, which are about 16 and 3.4 times larger than the minimum TBP of Eq. (1). This leaves room for further improvements. Ideal filtering will permit a distance of 50 km for 0 dBm launch power and 90 km for adapted 10 Gb/s laser power.

Our results supersede a recent theoretical analysis, which purported to demonstrate the infeasibility, at any distance, of single photon detection based QKD in a DWDM network with a single 0 dBm data channel.[28] This analysis has led to a proposal, followed by some preliminary research,[29] to use continuous-variable (CV) systems instead, despite the fact that the secure key rates offered by CV-QKD will be orders of magnitude lower.[30] We point out that the pessimistic result obtained in Ref. [28] is just the result of conservative parameter choice. The choice of $\Delta t = 1$ ns and $\Delta \nu = 75$ GHz results in a TBP of 75, far exceeding the value obtained using realistic components, as well as the ultimate limit imposed by Eq. (1).

In conclusion, we have shown that single-photon based QKD systems can be deployed in a DWDM environment. In these systems, temporal filtering is intrinsic to the single photon detectors and spectral filtering is achieved with low-cost off-the-the-shelf telecom components. Our QKD system allows a secure key rate of 2.38 Mb/s over 35 km fiber and a maximum distance reach of 70 km, in the presence of error-free bidirectional 10 Gb/s data. We have also demonstrated high bit rate QKD with the standard data laser power of 0 dBm. These positive results show the potential of high speed QKD for securing future communication infrastructures.

K. A. Patel acknowledges personal support via the EPSRC funded CDT in Photonics System Development.

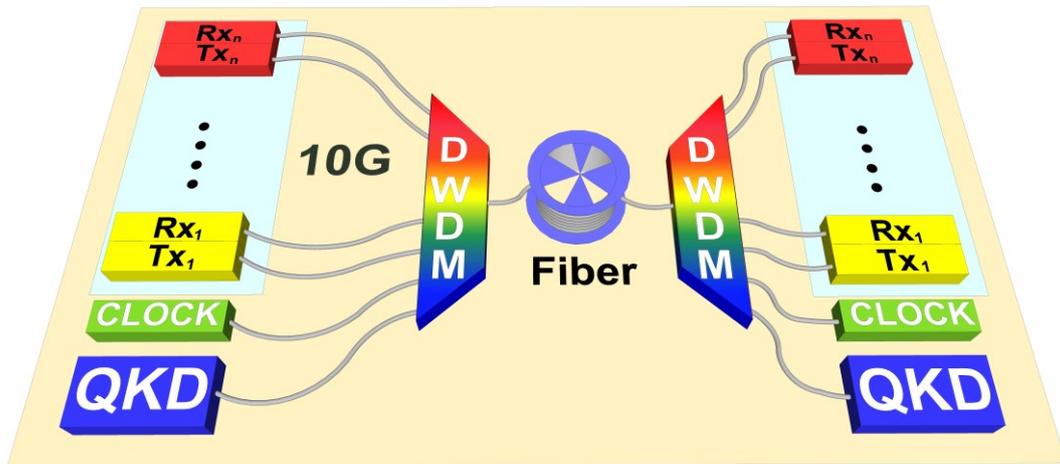

FIG. 1: QKD and 10 G data channel multiplexing schematic. DWDM (Dense wavelength division multiplexer), quantum and clock channels, pair of data channels 1,2…...n. Fiber spool length varies between 25 to 70 km.

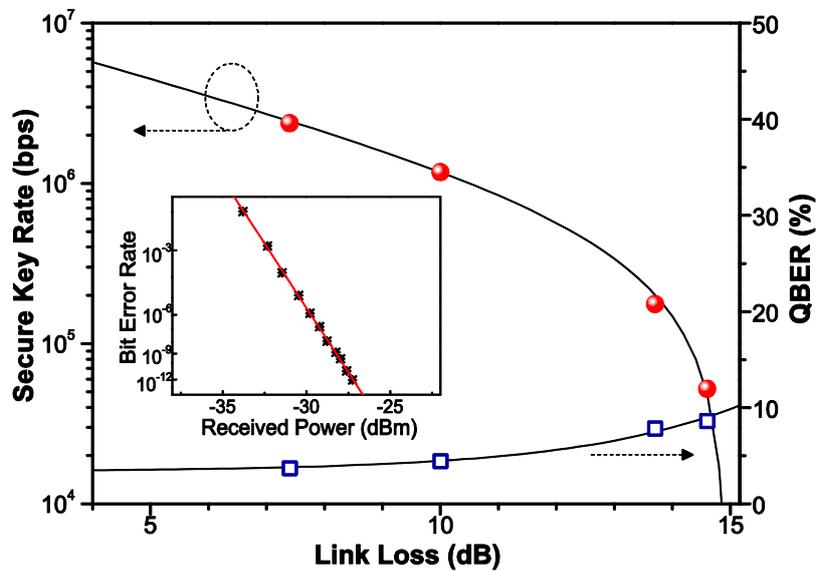

FIG. 2: Measured (symbols) and calculated (lines) QKD result as a function of fiber link loss. Inset: bit error rate versus received power for typical 10 G receiver.

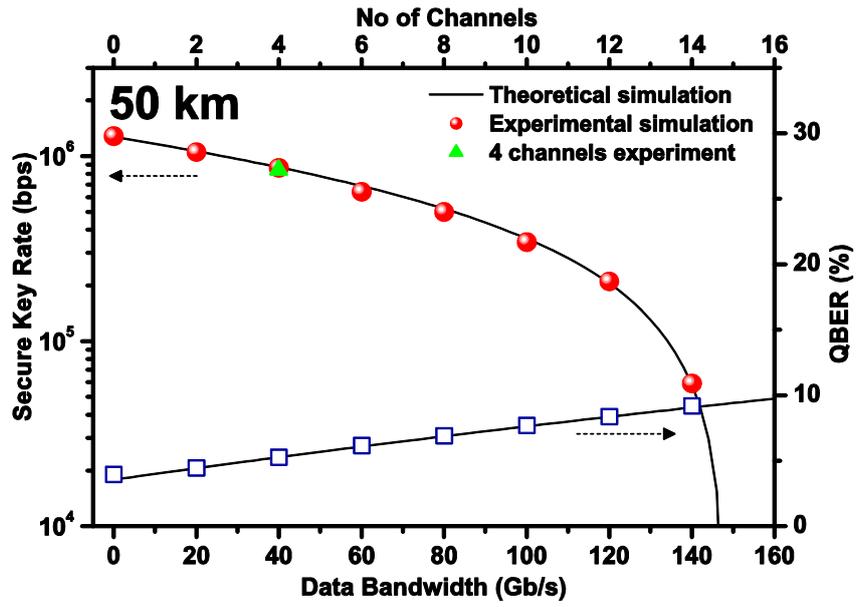

FIG. 3: Secure key rate and QBER versus data bandwidth at 50 km. The launch power for each simulated 10G channel is -14dBm.

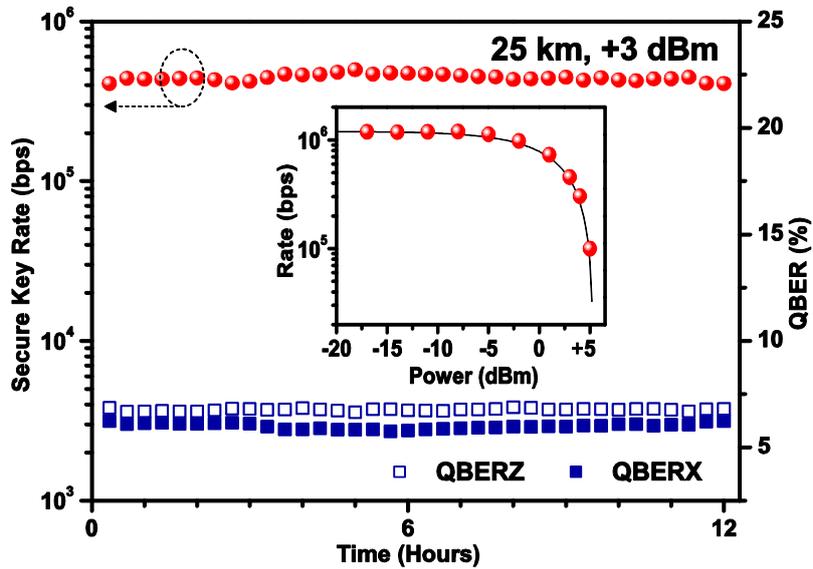

FIG. 4: 12 hours operation of QKD at 25 km with + 3 dBm classical launch power. Inset: experimental (symbols) and simulated (line) secure key rate *vs.* combined data laser power.